\documentclass[10pt,letterpaper,twocolumn]{article} %% two column, final layout

\usepackage{ol2}
\usepackage[draft]{hyperref}
\usepackage{amsmath}

\begin{document}

\twocolumn[ %% activate for two-column option

\title{Electro-optic adiabatic wavelength shifting and $Q$ switching demonstrated using a \textit{p-i-n} integrated photonic crystal nanocavity}

\author{Takasumi Tanabe,$^{1,2,3,*}$ Eiichi Kuramochi,$^{1,2}$ Hideaki Taniyama,$^{1,2}$ and Masaya Notomi$^{1,2,*}$}

\address{
$^1$NTT Basic Research Laboratories, NTT Corporation, 3-1, Morinosato Wakamiya, Atsugi, Kanagawa, 243-0198, Japan\\
$^2$JST-CREST, 4-1-8 Honmachi, Kawaguchi, Saitama, 332-0012, Japan\\
$^3$Currently with the Department of Electronics and Electrical Engineering, Keio University,\\
3-14-1 Hiyoshi Kohoku-ku, Yokohama, Kanagawa, 223-8522, Japan\\
$^*$Corresponding authors: takasumi@elec.keio.ac.jp, notomi@nttbrl.jp
}

\begin{abstract}
We demonstrate adiabatic wavelength shifting by electro-optic modulation using a \textit{p-i-n} integrated high-$Q$ photonic crystal nanocavity.  The wavelength of the trapped light is adiabatically shifted by modulating the resonance of the cavity faster than the photon lifetime.  The cavity resonance is changed by injecting electrons through a \textit{p-i-n} junction to reduce the refractive index.  In addition we employ adiabatic wavelength shifting in a demonstration of dynamic $Q$ tuning by electro-optic modulation.
\end{abstract}


] %% activate for two-column option

\noindent
A high-$Q$ photonic crystal (PhC) nanocavity can trap light for a very long time in a tiny space.\cite{Tanabe2007np, Noda2007}  Photons generally travel much faster than electrons, which makes the direct manipulation of photons very difficult.  When we can trap light in a small cavity for a period longer than the electrical modulation speed, we may be able to directly manipulate light properties, such as the wavelength.  Recently, a new concept for changing the wavelength of light, which we call adiabatic wavelength shifting, was proposed \cite{Notomi06} and demonstrated \cite{preble07, McCutcheon07, TanabePRL09} using high-$Q$ micro and nanocavities.  Adiabatic wavelength shifting can be explained by analogy with a guitar string.  When we pluck the string and then change its tension, the pitch will change.  The same phenomenon occurs with photons.  When we trap photons in an ultrahigh-$Q$ cavity and rapidly change its resonance, the wavelength of the trapped light should follow the cavity resonance because the photons cannot escape.  As a result, the wavelength of the trapped light will change.  This type of classical wavelength conversion has not been considered for optics because it was previously impossible to change the resonant wavelength of a small cavity within the photon lifetime.  Proof-of-principle demonstrations of adiabatic wavelength shifting have already been performed by pumping ultrahigh-$Q$ cavities from the top of a silicon (Si) slab to change the cavity resonance by employing the carrier plasma dispersion effect.

 In this study we demonstrate adiabatic wavelength shifting in an ultrahigh-$Q$ cavity system, but by using electro-optic (EO) modulation.  EO modulation is performed by integrating a \textit{p-i-n} junction with the PhC nanocavity.\cite{Tanabe2009oe, Tanabe2010apl}  We also discuss and demonstrate EO $Q$ switch with smallest size as an application of adiabatic wavelength shifting.  Although $Q$ switching has been realized by optical pumping,\cite{TanabePRL09, Xu2007, Tanaka2007} the demonstration of  a\textit{p-i-n} integrated EO $Q$-switch \cite{Manipatruni} is essential for the future development of integrated photonic memories on chips.

\begin{figure}[t]
 \begin{center}
  \includegraphics[width=3.5in]{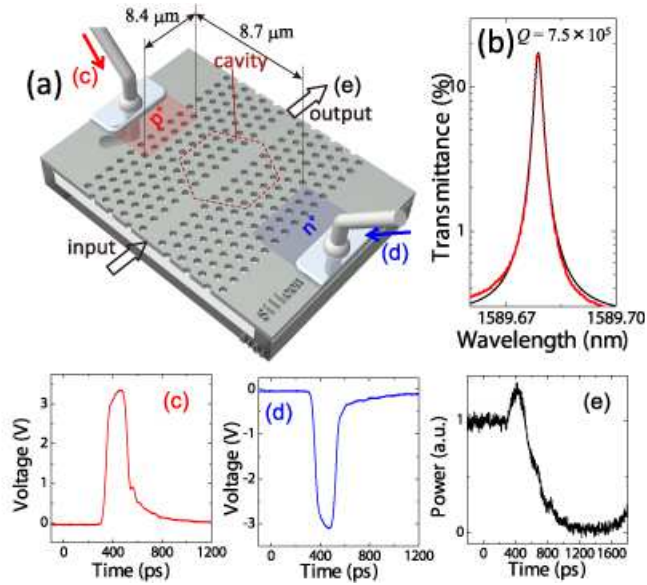}
  \caption{(Color online)
  (a) Schematic image of \textit{p-i-n} integrated Si PhC nanocavity.  The parameters of PhC nanocavity are described in Ref.~\onlinecite{Tanabe2010apl}.  (b) Measured transmittance spectrum.  (c) and (d) are the input electrical signal applied to the \textit{p} and \textit{n} junctions, respectively.  (e) is the EO modulated optical signal.
  }
  \label{fig1}
 \end{center}
\end{figure}

Figure~\ref{fig1}(a) is a schematic illustration of a fabricated \textit{p-i-n} integrated PhC nanocavity.  The fabrication process is detailed elsewhere.\cite{Tanabe2009oe} The PhC nanocavity consists of a width-modulated line defect,\cite{Kuramochi2006} and it is coupled to the input and output waveguides through barrier line defects that have a width of $0.98a\sqrt{3}$ (W0.98).\cite{Kuramochi2008}  The input and output waveguides are slightly wider at W1.05.  Figure~\ref{fig1}(b) shows the measured transmittance spectrum.  It exhibits a $Q$ of $7.5\times 10^5$, which corresponds to a photon lifetime of 0.63~ns.  Figures~\ref{fig1}(c) and (d) are the electrical signals that we employed for our experiments.  The positive voltage signal is applied to the \textit{p} junction and negative one to \textit{n} junction. (Note that both Fig.~\ref{fig1}(c) and (d) are forward bias signals.)  The current is 2.8~$\mathrm{\mu A}$ when total 6~V forward voltage is applied between \textit{p} and \textit{n} junctions.  Figure~\ref{fig1}(e) shows the EO modulated output optical waveform when a continuous-wave (CW) laser is applied at the cavity resonance. It shows that the EO modulation speed is about $\sim 0.5$~ns, which is about the same as the photon lifetime of the nanocavity, and this makes our demonstration possible.  The small peak that appears at $\sim 0.4$~ns results from the EO $Q$ switching that we describe later.

 First, we performed a wavelength resolved photon lifetime measurement.  It was basically the same as the measurement that we described in a previous study.\cite{TanabePRL09}  Briefly, we input a CW laser light that was in resonance with the PhC nanocavity (1589.78~nm) to charge the cavity with photons.  Then the input light was suddenly turned off at $t=0$~ps to enable us to observe the emission of the caged photons from the PhC nanocavity.  The input power level is kept very low in order to prevent the cavity exibiting any nonlinear effect.  The output power exhibited an exponential decay, and this gave $\tau_{\mathrm{ph}}$.  We placed a monochromator with a wavelength resolution of 0.04~nm in front of the point at which we measured the waveform.  A two-dimensional image was obtained by changing the wavelength of the monochromator after each waveform measurement.

\begin{figure}[tb]
 \begin{center}
  \includegraphics[width=2.8in]{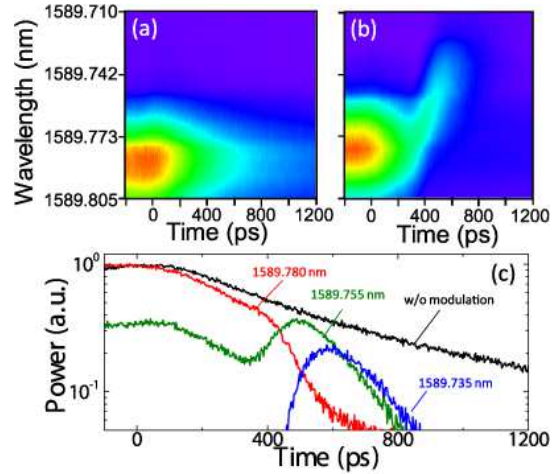}
  \caption{(Color online)
  (a)~Output spectrogram of the output light from the PhC nanocavity when no modulation is applied.
  (b)~Output spectrogram when the cavity is modulated.  A forward bias signal (shown in Fig.~\ref{fig1}(c) and (d)) is applied at $t\simeq 300$~ps. (c) Spectrally resolved waveforms obtained by taking line plots along the temporal axes of (a) and (b).   The black line indicates 1589.785~nm in (a), and other lines are taken from (b).
  }
  \label{fig2}
 \end{center}
\end{figure}

The measured spectrograms from the PhC nanocavity are shown in Fig.~\ref{fig2}.  When no modulation was applied the output light contained only the original input wavelength as shown in Fig.~\ref{fig2}(a).  Now we applied a forward bias voltage to the \textit{p-i-n} junction to inject electrons and holes.  This reduced the refractive index of Si as a result of the carrier plasma dispersion effect and the cavity resonance was modulated to a shorter wavelength.  Therefore, when we apply a forward bias during the cavity decay, we should observe adiabatic wavelength shifting.  Figure~\ref{fig2}(b) shows the output spectrogram when EO modulation was applied at $t\simeq 300$~ps.  The generation of a short wavelength component is clearly visible in the spectrogram. This light wavelength did not enter the PhC nanocavity.  We could also resolve the adiabatic transition of the light wavelength in time because the photon lifetime and the modulation speed are about the same.  (In contrast, we need to inject carriers much faster than the photon lifetime if we want to achieve larger wavelength shift.)  Figure~\ref{fig2}(b) shows that the cavity maintain single mode throughout the process and the wavelength is shifting continuously.  All these are direct evidences of adiabatic wavelength shifting.\cite{TanabePRL09}

To study the phenomena in more detail, we obtained a cross-sectional image along the temporal axis of the spectrograms in Figs~\ref{fig2}(a) and (b), and plotted them in Fig.~\ref{fig2}(c).  When the cavity is not modulated, the output waveform exhibits a smooth exponential decay that gives the original photon lifetime.   On the other hand, when the cavity is modulated, the output waveform of the original wavelength exhibits a sharp decay and a shorter wavelength component starts to appear.  The clear generation of a short wavelength component is confirmed.

It should be noted that adiabatic wavelength shifting can convert a light wavelength without it losing its coherence, because the light is not converted into electrons during the shifting process.  In addition, unlike the conventional nonlinear wavelength conversion process, the conversion efficiency of the adiabatic wavelength shifting does not depend on the trapped light intensity, and it can be performed on a single photon.  These will be powerful features when we attempt to use this device operation for quantum optics applications.

Next, we discuss the dynamic tuning of the $Q$ of an ultrasmall PhC nanocavity by EO modulation.  Although a high-$Q$ cavity can store photons for a long time, the high-$Q$ will limit its speed.  We need a high-$Q$ for storing photons, but we need to switch to a low-$Q$ mode for the faster charging and release of the optical pulses.  A number of groups, including ours, have already reported the successful demonstration of the dynamic tuning of the $Q$ on a Si chip.\cite{TanabePRL09, Xu2007, Tanaka2007}  However, in these studies the modulation is performed by optical pumping from the top of the slab, and such a scheme does not allow chip integration.  As far as we know, the \textit{p-i-n} PhC nanocavity is the smallest EO $Q$ switch cavity.  Therefore, a demonstration of the EO $Q$ switching of an ultrasmall PhC nanocavity will be an important step towards to the development of a chip integrated photon memory.

Previously we demonstrated the release of a short pulse from an ultrahigh-$Q$ PhC nanocavity as an application of adiabatic wavelength shifting.\cite{TanabePRL09}  The operating principle described in this section is the same except that we employ electrical injection instead of optical pumping.  We know that the frequency modegap of the W0.98 line defect lies at a higher frequency than that of the width modulated line-defect cavity.  This enables the strong confinement of photons in the cavity (high-$Q$ mode), because the cavity is well isolated from the waveguides.  When we shorten the wavelength (higher frequency) of the trapped light by adiabatic wavelength shifting, the cavity mode becomes closer to the mode-gap frequency of the W0.98 line defect.  This strengthens the coupling between the cavity and the waveguides, resulting in a low-$Q$ mode.

To demonstrate $Q$ switching, we performed the same experiment as in Fig.~\ref{fig2} but we fed the output waveform directly into the optical waveform measurement apparatus.
\begin{figure}[tb]
 \begin{center}
  \includegraphics[width=2.1in]{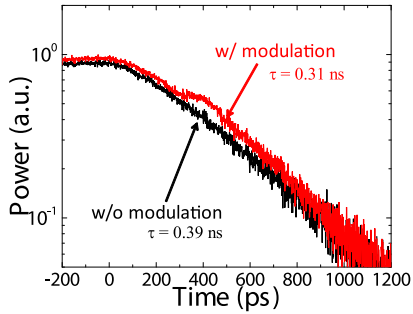}
  \caption{(Color online) EO $Q$ switching experiment.  Forward bias (Fig.~\ref{fig1}(a)) is applied at $t\simeq300$~ps. %Since the measurement is performed by repetitive modulation, the residual background carriers reduce the $Q$ of the cavity (which can be improved by improving the fabrication of the \textit{p-i-n} junction).
  To exclude the effect of residual background carriers, the black line is measured by modulating the cavity using a timing far from the recording window.}
  \label{fig3}
 \end{center}
\end{figure}
Figure~\ref{fig3} shows the recorded waveform with and without EO modulation.  When we modulated the device we observed a small increase in the output light, which is the result of $Q$ switching.  It shows that the coupling between the cavity and the output waveguide became stronger due to the adiabatic wavelength shifting.  The pulse readout speed to the output waveguide became also faster.  Note that the modulation of the photon lifetime is resulted by the modulation of the waveguide-cavity coupling and by free carrier absorption (FCA).%  According to the numerical analysis the $Q$ will change about 10\% when the index is modulated 0.01\%\cite{TanabePRL09}.

Although the $Q$-switching contrast obtained in Fig.~\ref{fig3} is currently not very large, it can be improved by increasing the EO modulation speed.  We found that \textit{i} region was too wide and the resistance of the \textit{n} junction was too high (due to imperfect doping condition) for achieving high speed.  Let us assume that we want to obtain a memory contrast time of 1 byte (8 bits).  Since we need to switch the $Q$ faster than the photon lifetime of the low-$Q$ mode, we need a modulation speed that is at least eight times faster than the photon lifetime in the high-$Q$ mode.  A speed higher than 10~GHz has been achieved in Si microring \textit{p-i-n} modulators,\cite{Xu, Liu} and a higher $Q$ has been achieved in PhC nanocavities,\cite{Tanabe2007np, Noda2007} thus indicating that a memory contrast of 1 byte is within the scope of state-of-the-art technology.

Finally, we should emphasize the advantage of EO $Q$ switching over optical pumping\cite{TanabePRL09, Xu2007, Tanaka2007}.  It is difficult to create an ultrahigh-$Q$ mode with an optical pumping scheme.  This is because the generated carriers induce FCA (which reduces the $Q$).  We need to employ a \textit{p-i-n} junction and reverse biasing to extract the carriers.  Then we can switch the cavity to a low-$Q$ mode by forward biasing the \textit{p-i-n} (carrier injection), and switch to an ultrahigh-$Q$ mode by reverse biasing (carrier extraction) at an arbitrary timing.  Neither the high nor the low $Q$ mode is significantly affected by FCA.  To demonstrate reverse biasing, we need to improve the operating speed of our \textit{p-i-n}, which is now under investigation.

In summary, we demonstrated adiabatic wavelength shifting by EO modulating a \textit{p-i-n} integrated high-$Q$ Si PhC nanocavity.  We clearly observed the phenomenon by monitoring the wavelength of the trapped light.  We also demonstrated EO $Q$ switching as an application of adiabatic wavelength shifting, which is essential if we want to develop photonic memories on Si chips.  

We thank Dr. T. Tamamura, Dr. K. Nishiguchi, and Mr. D. Takagi for helping with the fabrication.

\end{document}